\documentclass[aps,prl,reprint,twocolumn,showpacs,floatfix,superscriptaddress]{revtex4-1}

\usepackage{amssymb,amsmath,amstext}                
\usepackage{graphicx}                                               
\usepackage{epstopdf}                                               
\usepackage{color}                                                     
\usepackage{bm}                                                        
\usepackage{appendix}                                              
\usepackage[utf8]{inputenc}
\usepackage{latexsym}
\usepackage{xcolor}
\usepackage{braket}
\usepackage{ulem}
\usepackage{siunitx}
\usepackage{umoline}
\usepackage{epsfig}
\usepackage{graphics}
\usepackage{amssymb}
\usepackage{psfrag}
\usepackage{mathtools}
\usepackage[english]{babel}
\usepackage[T1]{fontenc}
\usepackage{lmodern}
\usepackage{booktabs}

\usepackage{rotating}
\usepackage{multirow}

\definecolor{lblue} {RGB}{51,71,158}
\usepackage[colorlinks=true,citecolor=blue,linkcolor=blue,urlcolor=lblue]{hyperref}

\begin{document}

\title{Constraints induced delocalization}

 \author{Piotr Sierant}\email{psierant@ictp.it} 
\affiliation{The Abdus Salam International Center for Theoretical Physics, Strada Costiera 11, 34151, Trieste, Italy}
 \affiliation{Institute of Theoretical Physics, Jagiellonian University in Krakow, \L{}ojasiewicza 11, 30-348 Krak\'ow, Poland }

\author {Eduardo Gonzalez Lazo}
 \affiliation{The Abdus Salam International Center for Theoretical Physics, Strada Costiera 11, 34151, Trieste, Italy}
 \affiliation{SISSA, via Bonomea, 265, 34136 Trieste, Italy}

\author{Marcello Dalmonte}
\affiliation{The Abdus Salam International Center for Theoretical Physics, Strada Costiera 11, 34151, Trieste, Italy}
\affiliation{SISSA, via Bonomea, 265, 34136 Trieste, Italy}

\author{Antonello Scardicchio}
\affiliation{The Abdus Salam International Center for Theoretical Physics, Strada Costiera 11, 34151, Trieste, Italy}
\affiliation{INFN Sezione di Trieste, Via Valerio 2, 34127 Trieste, Italy}
 \author{Jakub Zakrzewski}
 \affiliation{Institute of Theoretical Physics, Jagiellonian University in Krakow, \L{}ojasiewicza 11, 30-348 Krak\'ow, Poland }
 \affiliation{Mark Kac Complex Systems Research Center, Jagiellonian University in Krakow, \L{}ojasiewicza 11, 30-348 Krak\'ow,
 Poland. }

\date{\today}

\begin{abstract}
We study the impact of quenched disorder on the dynamics of locally constrained
quantum spin chains, that describe 1D arrays of Rydberg atoms in both frozen 
(Ising-type) and dressed (XY-type) regime. 
Performing large-scale numerical experiments, we observe no trace of many-body 
localization even at large disorder. Analyzing the role of quenched disorder terms in 
constrained systems we show that they act in two, distinct and competing ways: as an 
on-site disorder term for the basic excitations of the system, and as an interaction
between excitations. The two contributions are of the same order, and as they compete 
(one towards localization, the other against it), one does never enter a truly strong 
disorder, weak interaction limit, where many-body localization occurs. Such a mechanism 
is further clarified in the case of XY-type constrained models: there, a term which 
would represent a \emph{bona-fide} local quenched disorder term acting on the excitations
of the clean model must be written as a series of non-local terms in the unconstrained 
variables. Our observations provide a simple picture to interpret the role of quenched 
disorder that could be immediately extended to other constrained models or quenched
gauge theories.

\end{abstract}

\maketitle

{\it Introduction.} 
The foundational hypothesis of statistical mechanics is that an isolated, quantum system will 
reach an equilibrium independent of the initial conditions, except for a few variables 
related to macroscopic conserved observables (i.e.\ particle number, total energy,
total momentum etc.). In a modern language, to describe the equilibrium of isolated systems, 
the role of the ergodic hypothesis of Boltzmann \cite{Gallavotti95} is taken by
the eigenstate thermalization hypothesis \cite{Deutsch91, Srednicki94, Alessio16}.
 
Recently, a generic mechanism that inhibits the approach to equilibrium of interacting quantum many-body systems
in the presence of disorder has been identified in the phenomenon of many-body localization (MBL) 
\cite{Basko06,Gornyi05,Znidaric08,Pal10}. In this dynamical phase, the information
about the initial state is preserved indefinitely in the values of local 
integrals of motion \cite{Huse14,Ros15}, the transport is 
suppressed \cite{Nandkishore15, Znidaric16,Alet18, Abanin19},
and the entanglement spreads slowly \cite{serbyn2013universal,iemini2016signatures}.
The original works on spinless fermions in any dimensions \cite{Basko06,Gornyi05} were
confirmed and supplemented by numerics on spin-1/2 XXZ chains \cite{DeLuca13,Luitz15}.
 The results for the disordered XXZ spin chains were followed by the study of bosonic models \cite{Sierant18, Orell19},
 systems of spinful fermions \cite{Mondaini15, Prelovsek16, Zakrzewski18, Kozarzewski18} or models with random interactions
 \cite{Sierant17, Lev16, Li17a} suggesting that sufficiently strongly disordered isolated quantum many-body
 systems always undergo MBL.
 
However, locating and characterizing the MBL transition is notoriously difficult. 
For the most studied system which shows MBL, the disordered XXZ spin chain, the 
maximal system size accessible to present day supercomputers for an exact 
treatment is $L=24$ \cite{Mace18, Pietracaprina18, Sierant20p, Beeumen20}.
On top of that, the observed
finite-size scaling is extremely slow. 
The ensuing difficulties in extrapolating results to the thermodynamic limit 
sparkled a recent debate about the existence of the MBL phase 
\cite{Suntajs19, Sierant20b, Abanin19a, Panda19} and its dynamical
properties \cite{Kiefer20, Luitz20, Sels20}. In parallel to these 
theoretical efforts, experiments in the synthetic quantum matter
have already probed regimes of strong interactions and strong disorder
necessary for MBL in both Ising- and Hubbard-type models \cite{Schreiber15,Luschen17}. 
Very recently, a new generation of platforms based on Rydberg atoms in optical lattices 
and tweezers~\cite{Bernien17,Barredo18,Zeiher17, Browaeys20, Guardado20}  has demonstrated 
an impressive capability to perform coherent dynamics up to considerably long timescales, 
allowing, for instance, for the realization of mesoscopic-sized ordered states~\cite{Scholl:aa,Ebadi:aa}.
Strong nearest-neighbor interactions that characterize these systems naturally lead to effective 
constrained dynamics in both Ising- and XY-type regimes. A natural question to ask is thus, 
whether the interplay between constraints, interactions and disorder can lead to a scenario
that is qualitatively different from the unconstrained models, and whether such a scenario 
can be characterized by common, generic features.

In this work, we show that 1D spin chains with local constraints can remain
ergodic even in the presence of a strong disorder. 
Such models are experimentally realized in arrays of ultracold Rydberg atoms 
\cite{Bernien17,Barredo18,Zeiher17, Browaeys20, Guardado20}. 
The local constraints arise in the Rydberg blockade regime and alleviate the 
exponential growth of Hilbert space with the  system size $L$.
This feature allows us to overcome the limitation of small system 
sizes that impede studies of the MBL transition in the conventional, unconstrained 
spin chains and to consider constrained models of sizes exceeding 
$L=100$ 
sites for the largest constraint radius considered.

Investigating the crossover between ergodic and MBL regimes, we see no signs of localization in the thermodynamic limit. 
We identify the reason for this behavior in a non-trivial action of a generic ``quenched disorder" term in a constrained model. 
Such a term does not simply act as an on-site disorder on the basic excitations of the clean system, but generically 
introduces an interaction term between them: in a representation in which the basic degrees of freedom of the system 
are \emph{unconstrained}, the quenched disorder term is written as both a random on-site energy term and a random 
denstity-density interaction. Both terms are of the same order and their interplay does not allow the system to 
be in a strong disorder, weak interaction regime in which MBL can be established in a controlled
manner~\cite{Basko06}. While we will be focusing on models that are motivated by the aforementioned 
experiments, we believe that this observation extends to generic constrained models.

{\it Disordered PXP models. } We consider a 1D chain of Rydberg atoms 
in the frozen regime and assume that strong interactions allow the excitation to Rydberg state only for pairs of atoms separated
by at least $\alpha$ sites~\cite{Lesanovsky12,Bernien17,Barredo18,Zeiher17, Browaeys20}.
This leads to the Hamiltonian
\begin{equation}
 \hat{H}= \sum_{i=1}^{L} P^{\alpha}_{i} S^x_i P^{\alpha}_{i+1+\alpha}+
\sum_{i=1}^{L} h_i S^z_i 
 \label{Hpxp}
\end{equation}
where the projectors $P^{\alpha}_i=\prod_{j=i-\alpha}^{i-1}(1/2-S^z_j)$ 
assure that the dynamics is confined to a constrained Hilbert space, $h_i$ are independent, uniformly distributed
random variables in the interval $[-W/2,W/2]$ with $W$ being the disorder strength 
and with periodic boundary conditions (PBC) $\vec{S}_{L+i}\equiv \vec{S}_{i}$ assumed.

The clean ($h_i=0$) PXP models are known to host many-body 
scar states for a constraint radius $\alpha=1$ \cite{Turner18,Lin19, Wen19, Khemani19, Choi19, Ho19, Iadecola19, Iadecola19a,Surace19}
as well as for $\alpha \geq 1$ \cite{Surace20} and even in presence of disorder \cite{Mondragonshem20}.
The scar states are, however, not statistically important
for the properties of generic eigenstates that are of direct interest here. 
On the other hand, for a PXP model with disorder on both the $S^x$ and $S^z$ terms,  both an ergodic and an MBL regimes were claimed to exist  \cite{Chen18pxp}. That was interpreted in favour of a stable MBL phase at large disorder strengths. 
For the blockade radius $\alpha$, the Hilbert space dimension is $\mathcal N_{\alpha} =(\Phi_{\alpha})^L$ 
where $\Phi_{\alpha}\approx 1.6180, 1.4656, 1.2852$
respectively for $\alpha=1,2,5$ \footnote{$\Phi_\alpha$ solves the equation $\Phi^{\alpha+1}-\Phi^\alpha-1=0.$ 
Although there is no explicit solution of this equation for $\alpha>3$ one finds the
asymptotic form $\Phi_\alpha=1+\frac{W(\alpha)}{\alpha}+...$ where $W$ is the Lambert 
$W$-function solving the equation $W(x)e^{W(x)}=x$. Already at $\alpha=5$ the error is about $1\%$. }.
This allows us to access progressively larger system sizes
with increasing $\alpha$ while studying 
the crossover between ergodic and MBL regimes.
Similar ideas, employing local constraints, 
were used to demonstrate a presence of MBL regime in 2D  dimer systems \cite{Theveniaut20,Pietracaprina20}
and to study MBL in Krylov spaces of a pair-hopping model \cite{Herviou20}.

 \begin{figure*}
 \includegraphics[width=0.99\linewidth]{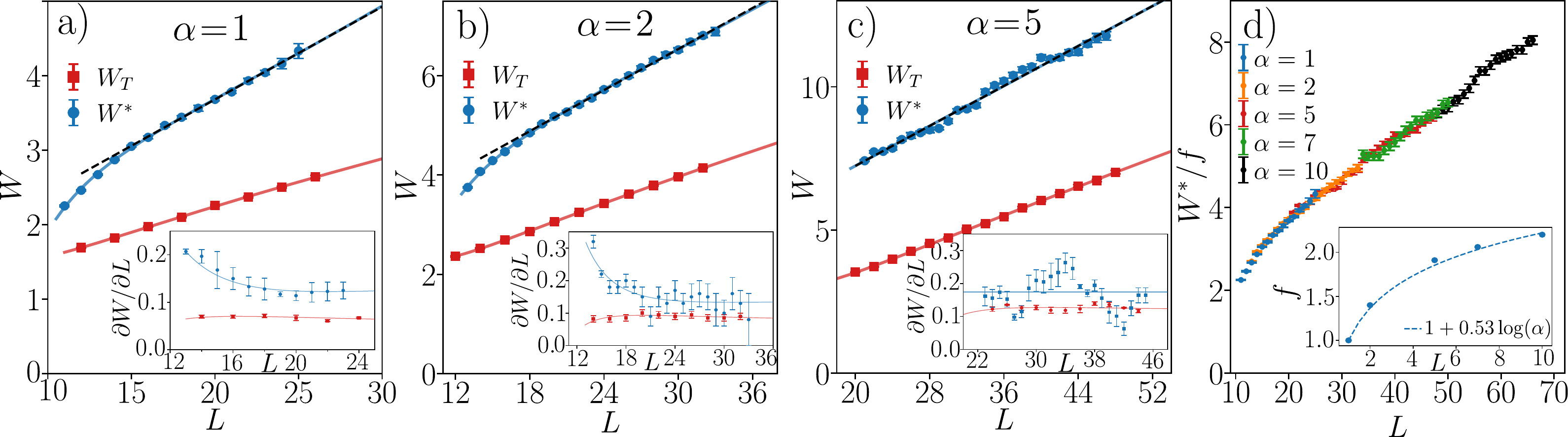} \vspace{-0.3cm}
  \caption{The ergodic-MBL crossover in disordered PXP \eqref{Hpxp} models. 
  Panels a)-c): disorder strengths $W_T$ and $W^*$ as function of system size $L$. Solid lines
  denote fits of the form $W(L)=aL+b+c/L+d/L^2$, dashed lines correspond to $W(L)=aL+b$. 
  The insets show derivatives of data with respect to $L$. Panel d) shows $W^*(L)$ rescaled
  by a factor $f$ for varying radius of constraint $\alpha$.
 }\label{figCOMBrbar}
\end{figure*}

{\it Ergodic-MBL crossover in PXP models. } We calculate eigenvalues $E_i$ and eigenstates of disordered PXP models
\eqref{Hpxp} for $\alpha=1,2,5$ using full exact diagonalization for system sizes $L$ for which the Hilbert space
dimension $\mathcal N_{\alpha} \leq 10^4$ and POLFED algorithm \cite{Sierant20p} for larger $L$.
We compute $r_i=\min \{g_{i},g_{i+1} \} / \max\{g_{i},g_{i+1}\} $ (where 
$g_i=E_{i+1}-E_{i}$), average it over 
$\min\{\mathcal N_{\alpha}/20,1000 \} $
of the eigenvalues 
from the middle of the spectrum and subsequently average the results over disorder realizations
to obtain the average gap ratio $\overline r$. The number of disorder realizations varies 
between a few millions for the smallest $L$ down to no less than $2000$ ($5000$) for the largest
(second largest) system size $L$ considered for a given model. The average gap ratio $\overline r$
reflects properties of level statistics changing between
$\overline r_{GOE}\approx 0.53$ for an ergodic system 
and $\overline r_{PS}\approx 0.386$ for a localized system \cite{Atas13}.
Indeed, we observe that for small disorder strengths $W$  
the average gap ratio in the considered models is $\overline r = \overline r_{GOE}$, and
that it decreases to $\overline r \approx \overline r_{PS}$ for large $W$, see \cite{suppl}.

To investigate the crossover between the ergodic and MBL regimes we introduce two 
system-size dependent disorder strengths: i) $W_T(L)$ -- the disorder strength for which, at
a given system size $L$, the average gap ratio starts to deviate from the ergodic value
and is equal to $r_{GOE}-p$ 
(we choose $p=0.01$ but other choices of $p<0.02$ lead to 
quantitatively similar results); 
ii) $W^*(L)$ -- the disorder strength for which curves $\overline r(W)$
cross for system sizes $L_1$ and $L_2$ such that $L=(L_1+L_2)/2$, we use 
$2\leq|L_1-L_2|\leq 4$ for $\alpha=1,2$ and 
$4\leq|L_1-L_2|$ for $\alpha \geq 5$ models. 

The resulting $W_T(L)$ and $W^*(L)$ curves
divide the phase diagram into three
regimes: ergodic for $W<W_T(L)$ with $\overline r(W)=r_{GOE}$; ``critical'' for $W_T(L)<W<W^*(L)$
in which the value of $\overline r(W)$ increases with system size $L$ towards 
$r_{GOE}$; MBL for $W>W^*(L)$ in which the average gap ratio $\overline r(W)$
decreases down to $\overline r_{PS}$ value with increasing $L$.

For the widely studied 
disordered XXZ model 
\cite{Santos04a, Oganesyan07, Berkelbach10,Luitz15,   Bera15, Enss17, Herviou19, Sierant20, Colmenarez19, Chanda20m},
one observes the scalings 
 $W_T(L)\sim L$ and $W^*(L)\sim W_C+c/L$ \cite{Sierant20p}.
Extrapolating the scaling $W^*(L)\sim W_C+c/L$ to $L\rightarrow \infty$, one gets a critical disorder
strength $W_C\approx5.4$, slightly larger than the usually cited value $W_C\approx 3.7$ \cite{Luitz15} 
but consistent with various lower bounds \cite{Devakul15, Gray18, Doggen18}.
At the same time, the two scalings  $W_T(L)\sim L$ and $W^*(L)\sim W_C+c/L$ become 
 incompatible for system sizes larger than $L_0\approx 50$ (a length scale which appeared before, for this model \cite{Panda19}).
 The asymptotic regime $L>L_0$ is well beyond reach of present day supercomputers for XXZ model, 
 hence evidence for either of the scalings to prevail in the thermodynamic limit is lacking. We show below that the situation is much clearer for disordered constrained models.

For disordered PXP models we observe a linear dependence 
$W_T(L)\sim L$ as shown in Fig.~\ref{figCOMBrbar} (a-c). The disorder strength $W^*(L)$, 
describing the drift of the crossing point with system size, 
shows a clear curvature at small $L$ suggesting the $W^*(L)\sim W_C+c/L$ scaling.
However, for $L\gtrapprox 20$ for $\alpha=1$ ($L\gtrapprox 22$ for $\alpha=2$)
this curvature vanishes and $W^*(L)$ starts to grow linearly with the system size $L$. 
see Fig.~\ref{figCOMBrbar}(a-b).
Importantly, for $\alpha=2$ the interval of system sizes for which the linear drift $W^*(L)\sim L$ is observed
is wider than for $\alpha=1$. Increasing the radius of the blockade further, to $\alpha=5$,
we still see -- Fig.~\ref{figCOMBrbar} c) -- a linear dependence $W_T(L)\sim L$. The curvature 
of the $W^*(L)$ curve, observed for smaller system sizes 
disappears for $\alpha=5$.
Instead, we observe a linear drift $W^*(L)$ with a small oscillation on top of it for all available system sizes.

These conclusions are further supported by the derivatives $\partial W_T/\partial L $, $\partial W^*/\partial L $ 
shown in the insets of Fig.~\ref{figCOMBrbar}. The derivative $\partial W_T/\partial L $ clearly approaches a constant
$w_T$
with the increasing system size. The derivative $\partial W^*/\partial L $ decreases
with $L$ for $\alpha=1,2$, oscillates around a constant for $\alpha=5$
and is bound from below by $w_T$. This is consistent the 
linear drift of the both disorder strengths $W_T(L)\sim w_T L $ and $W^*(L) \sim w^* L $ with $w_T\leq w^*$ at sufficiently large $L$.
The recent observation that disorder strength required for localization of wavefunctions in Fock space (which is a stronger 
ergodicity breaking than MBL) occurs at disorder strength $W'=w'L$ of \cite{DeTomasi20} is consistent with our
results provided $w' > w^*$. A similar analysis can be also performed for bipartite entanglement entropy of
 eigenstates with the same conclusions~\cite{suppl}.

 Superimposing results for various constraint radius $\alpha$ as shown in Fig.~\ref{figCOMBrbar} d), we observe that they fall on
 top of a universal curve if the crossing points $W^*(L)$ are rescaled by factor $f$ that increases
 approximately logarithmically with $\alpha$. This is surprising at first sight: one could expect that 
 the larger radius of constrain $\alpha$ implies smaller number of spin flips for a given spin configuration 
 enhancing localization in the system. In fact, exactly the opposite is true. To see that, consider a spin configuration 
 with maximal number of spins up for given $\alpha$ which is roughly equal to $L/(1+\alpha)$. To perform a transition to a spin configuration
 with all spins up shifted by one lattice site, it suffices to act $2L/(1+\alpha)$ times with the kinetic term of the Hamiltonian \eqref{Hpxp}.
 Hence, the kinetic term becomes more effective with increasing radius of constraint favoring delocalization at
 larger $\alpha$ (see also \cite{suppl}). However, this is not the ultimate reason of why we observe delocalization in the constrained models
 as we show below.
 
  \begin{figure*}
 \includegraphics[width=0.99\linewidth]{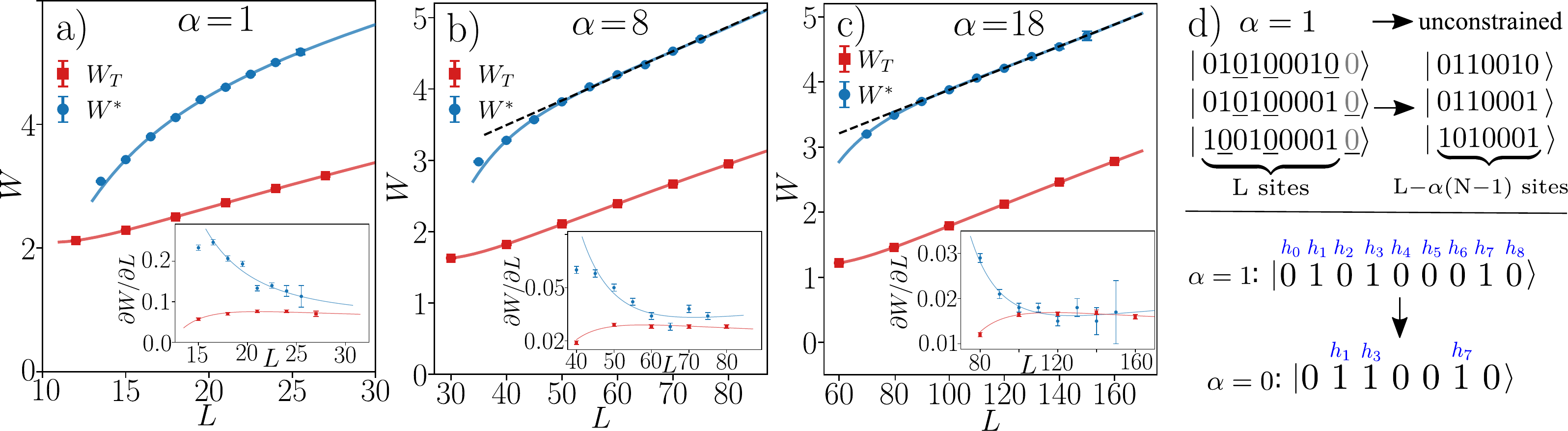} \vspace{-0.3cm}
  \caption{ The ergodic-MBL crossover in disordered constrained models with $U(1)$ symmetry \eqref{Hpxyp}. 
  Panels a)-c): disorder strengths $W_T$ and $W^*$ as function of system size $L$. Panel d) (top):
  the mapping between constrained and unconstrained models: $\alpha$ unoccupied sites
  added after the right end of the chain are denoted in gray, the underlined sites are joined to the 
  neighboring particles to form the particles of the unconstrained model. Panel d) (bottom)
  the potential felt by the 
  particles before and after the mapping.
 }\label{figCOMBrbar2}
\end{figure*}  

 \textit{A constrained model with $U(1)$ symmetry.}
 Consider a system of spinless fermions with Hamiltonian
 \begin{equation}
 H = \sum_{i=1}^L  P_i \left(  c^{\dag}_i c_{i+1} +  c^{\dag}_{i+1} c_i \right) P_{i+2+\alpha}+ \sum_{i=1}^L h_i n_{i}
 \label{Hpxyp}
\end{equation}
 where $c^{\dag}_i$ ($c_{i}$) are fermionic creation (annihilation) operators, the particle number operator is $n_i=c^{\dag}_i c_i$,
 the projectors $P^{\alpha}_i=\prod_{j=i-\alpha}^{i-1}(1-n_j)$ assure that the particles are separated by at least $\alpha$ sites,
  $h_i$ are independent, uniformly distributed random variables in the interval $[-W/2,W/2]$ and PBC are imposed.
 The model \eqref{Hpxyp} maps, via the Jordan-Wigner transformation, to model analogous to
 disordered PXP model \eqref{Hpxp} with $S^x_i$ replaced by a spin-flip term $S^x_i S^x_{i+1}+S^y_i S^y_{i+1} $ in
 the kinetic term. Hence, \eqref{Hpxyp} is a natural extension of the constrained model \eqref{Hpxp} that
 has the $U(1)$ symmetry and preserves the total magnetization $S^z=\sum_{i=1}^L S^z_i$ (or equivalently, 
 the total number of particles $N=\sum_{i=1}^L n_i$). We consider a filling $N/L=1/(\alpha+2)$.
 The model \eqref{Hpxyp} (for $\alpha=1$) can be viewed as a strong interaction limit of XXZ spin chain \cite{DeTomasi19} 
 and may be realized experimentally by Rydberg dressing technique~\cite{Pupillo:2010aa,Honer:2010aa,Henkel:2010aa, Mattioli:2013aa,Guardado20}.

In contrast to the PXP models, 
the kinetic term of the constrained model \eqref{Hpxyp} does not become more effective when the 
constraint radius $\alpha$ is increased. Indeed, due to the particle number conservation, roughly $L/2$ actions of the 
kinetic term are needed to reach an arbitrary Fock state from a given starting Fock state (see also \cite{suppl}).
Based on the argument above, one could then expect that the constrained models with $U(1)$ symmetry are much more prone to localization.
This, however, turns out not to be the case, as revealed by an analysis of the crossover in the
average gap ratio $\overline r$ between the ergodic and MBL regimes. 
The disorder strengths $W_T(L)$ and $W^*(L)$, shown in Fig.~\ref{figCOMBrbar2}~a)-c)
are similar to the results for the disordered PXP models. The boundary of the ergodic regime $W_T(L)$, behaves linearly 
in $L$, $W_T(L) \sim w_T L$. The boundary of the MBL regime, $W^*(L)$, shows some curvature at smaller $L$ but 
then approaches a linear behavior $W^*(L)\sim w^* L$, with slope $w^*\geq w_T$. Hence, the whole crossover between
ergodic and MBL regimes drifts linearly with $L$ towards increasing disorder strengths and the systems delocalize 
in the thermodynamic limit. 

{\it On-site disorder in presence of constraints. } To understand  the delocalization of the constrained models, let us reconsider 
the Hamiltonian \eqref{Hpxyp}, and assume open boundary conditions (OBC) for simplicity. The presence of constraints 
prevents the particles from approaching each other at a distance smaller than $\alpha$. Hence, it is possible to 
associate an \textit{excluded volume} of $\alpha$ sites with each of the particles (for instance to the right
of the particle). Then, by adding $\alpha$ unoccupied sites at the right end of the chain, one can replace each particle
and $\alpha$ sites to its right, by an occupied site of a new, smaller chain. This shrinking procedure, illustrated
in Fig.~\ref{figCOMBrbar2}~d), defines a one-to-one mapping between Fock states 
of system of $L$ sites with constraint of radius $\alpha$ and between Fock states of an \textit{unconstrained}
system of spinless fermions on $L-\alpha(N-1)$ sites. Moreover, the particles can hop in the same manner   
before and after the mapping (if a given particle cannot hop, say, to the right in the constrained model
due to a presence of another particle $\alpha$ sites to its right, it also cannot hop to the right in the unconstrained model 
since the neighboring site is occupied). This means that in the absence of disorder ($h_i=0$), the Hamiltonian of the model
\eqref{Hpxyp} for $N$ particles on $L$ sites with constraint radius $\alpha$ and OBC exactly coincides with a Hamiltonian 
of $N$ spinless fermions on $L-\alpha(N-1)$ sites. Thus, for $h_i=0$ the constrained model \eqref{Hpxyp}, which is a non-Gaussian
fermionic model can be mapped to a non-interacting system.

What happens when disorder is introduced to the system? The model \eqref{Hpxyp} becomes interacting \textit{due to}
the presence of on-site disorder term $\sum_i h_i n_i$. To see this, consider again the
mapping between constrained and unconstrained model, as shown in the bottom panel of
Fig.~\ref{figCOMBrbar2}~d). For the constrained model, a particle at site $i$ experiences the potential $h_i$. After the mapping,
the particle at site $i$ of the unconstrained model feels the potential $h_{i+\alpha N_i}$ (where $N_i\equiv \sum_{j=1}^{i-1}n_i$).
In that way the disorder becomes a source of interactions in the constrained model \eqref{Hpxyp} since the potential felt by 
a particle on site $i$ depends on the total number of particles to its right. Rewriting 
 $\sum_{i=1}^{L} h_{i+N_i} n_i = \sum_{i} h_i n_i D^0_i+\sum_{i} h_{i+\alpha} n_i D^1_i+\ldots$,
where $D^0_i=\prod_{j<i}(1-n_j)$ [$D^1_i=\sum_k n_k \prod_{j<i, j\neq k}(1-n_j)$] is non-zero if there is exactly 
$0$ [$1$] particles on sites $1\leq j <i$ and further terms contain analogous terms that are non-zero if there are
$1<n < N$ particles on sites $1\leq j <i$. Hence, the on-site disorder term introduces 
random interactions of infinite range to the model the constrained model is mapped to. Moreover, the strength of
 interactions is increasing
with disorder strength $W$. Those two factors are at the root of the numerically observed delocalization of
constrained models. Importantly, while our mapping between constrained and unconstrained models does not 
directly apply to disordered PXP models (since the varying number of spins up translates into varying length of the
unconstrained chain), the mechanism in which disorder in presence of constraints provides interactions in the system 
is at play also in those models.

{\it Conclusions. } Studying the crossover between ergodic and MBL regimes
in locally constrained quantum spin chains, we observe that 
the whole crossover shifts linearly to larger disorder strength $W$ with increasing 
system size. This trend, thanks to the availability of larger system sizes in the constrained Hilbert
space geometry,  is well documented and occurs both in the disordered PXP models as well as in models with
$U(1)$ symmetry.

We argue that the observed delocalization can be traced back to fundamentally different
roles played by the on-site disorder in conventional and constrained models.
A sufficiently strong disorder leads to MBL in the former models
as was exemplified for the disordered transverse field Ising model \cite{Imbrie16,Imbrie16a}.
In contrast, for the constrained models studied here, the disorder can be seen as giving raise not only to one-body terms that tend to localize the fundamental excitations of the clean system but necessarily also to interactions that become stronger when disorder strength is increased. Depending on the details of the model, these interactions can be sufficiently strong to ultimately lead to delocalization. A similar phenomenon is expected to occur in quenched gauge theories with non-trivial center \cite{Brenes18}. The family of disordered, constrained quantum spin chains models considered in this work can be investigated experimentally in Rydberg atom setups. From the theoretical perspective, it emphasizes the richness and potential generality of dynamics arising out of the competition between interactions, disorder, and constraints.

 {\it Acknowledgments. } 
We are grateful to Giuliano Giudici for contributions at early stages 
of this work and relevant discussions. We also acknowledge interesting discussion with 
Lev Vidmar.
The work of EGL and MD is partly supported by the ERC under grant number 758329 (AGEnTh), by
the Quantera programme QTFLAG, by the MIUR Programme FARE (MEPH), and has received
funding from the European Union's Horizon 2020 research and innovation programme
under grant agreement No 817482. This work has been carried out within the activities
of TQT. JZ acknowledges support of National Science Centre (Poland) via grants OPUS18 2019/35/B/ST2/00034 
and Unisono 2017/25/Z/ST2/03029 corresponding to the QTFLAG Quantera programme.
P.S. acknowledges the support of  Foundation  for
Polish   Science   (FNP)   through   scholarship   START.

\normalem
%



\newpage

\begin{figure*}
\includegraphics[width=1\linewidth]{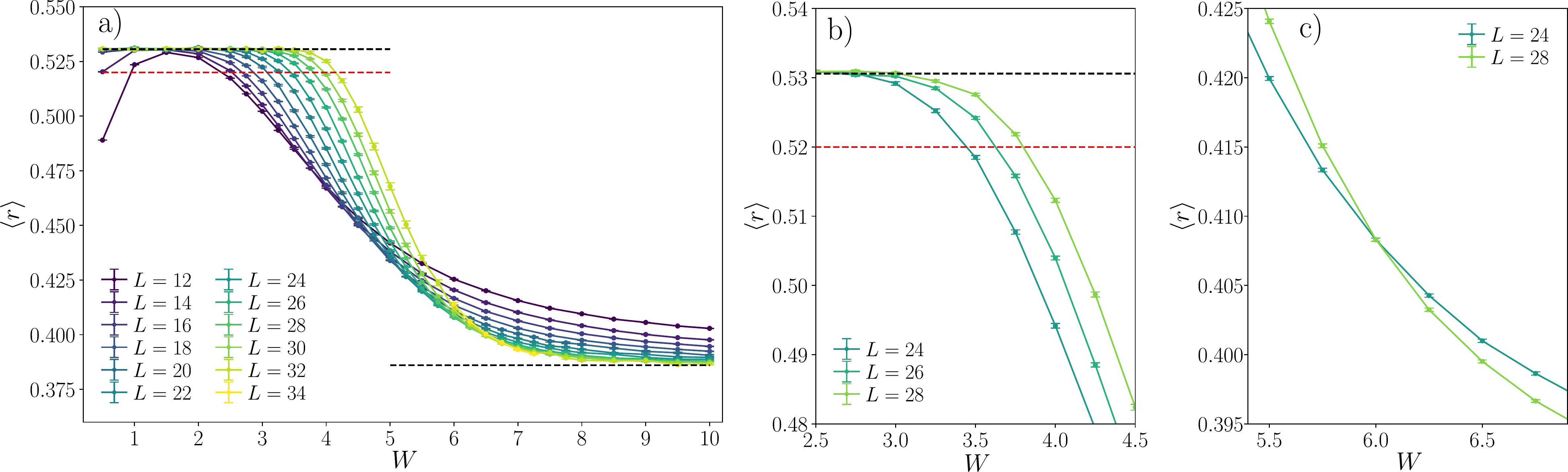}\vspace{-0.3cm}
  \caption{ Determination of disorder strengths $W_T(L)$ and $W^*(L)$.
  a) The average gap ratio at the middle of the spectrum for disordered PXP model with constraint radius $\alpha=2$;
  b) $W_T(L)$ is found as a disorder strength $W$ for which $\overline r(W) =r_{GOE}-p\approx 0.52$ (denoted by the red line);
  c) The crossing of the gap ratio vs disorder strength $r(W)$ curves for $L=24$ and $L=28$ determined $W^*(L=26)$.
 }\label{figS1}
\end{figure*} 

\begin{figure*}
\includegraphics[width=0.95\linewidth]{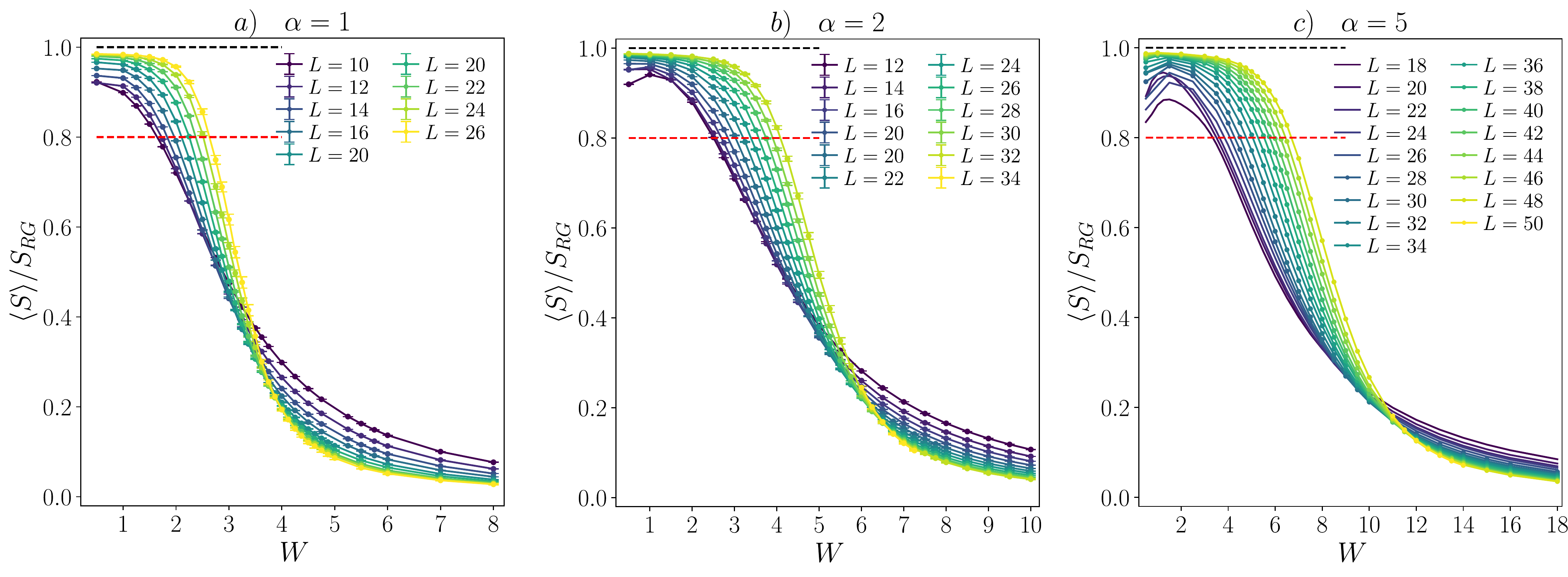}\vspace{-0.3cm}
  \caption{The average entanglement entropy  $\left \langle S  \right \rangle$ of eigenstates at the middle of the spectrum of disordered
  PXP models divided by the entanglement entropy of random Gaussian states $S_{RG}$ as a function of disorder strength $W$ for various 
  system sizes $L$ and radius of constraint $\alpha$.
 }\label{figEnt}
\end{figure*} 
\begin{figure*}
 \includegraphics[width=0.85\linewidth]{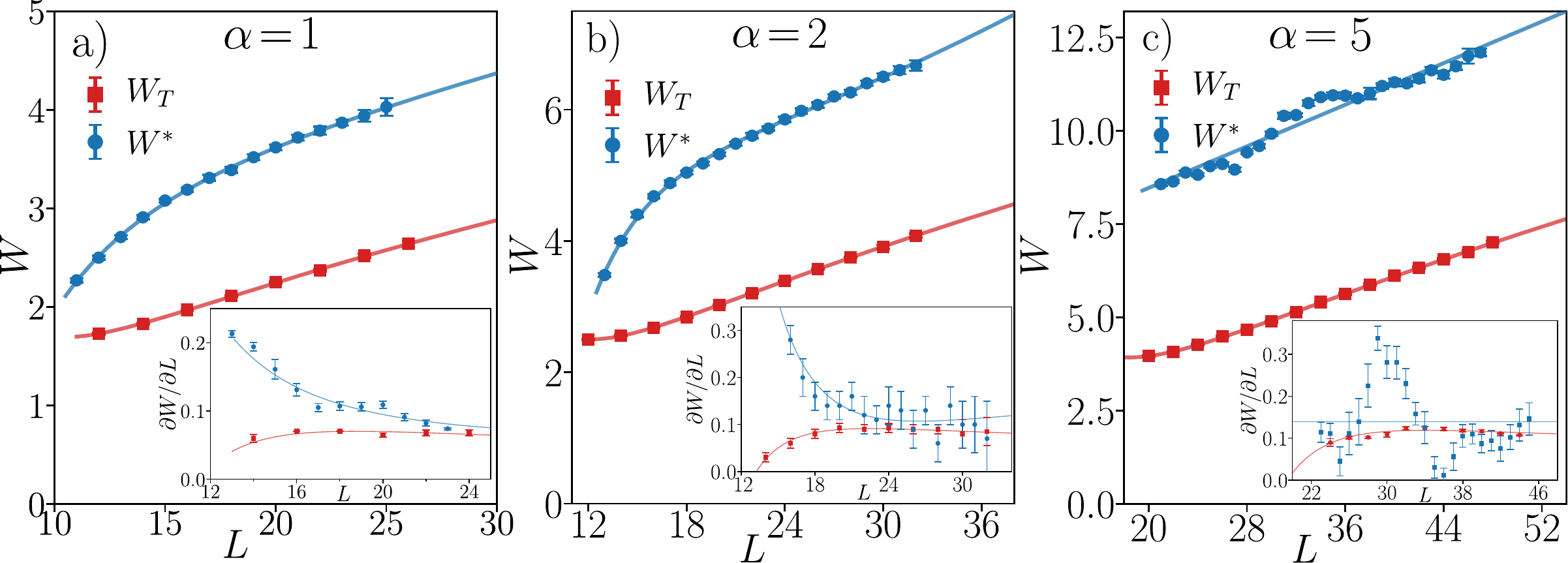} \vspace{-0.3cm}
  \caption{
  The ergodic-MBL crossover in disordered PXP \eqref{Hpxp} models. Disorder strengths $W_T(L)$ and 
  $W^*(L)$ are obtained from analysis of the rescaled average entanglement entropy of eigenstates
  and plotted as function of the systems size $L$. Solid lines
  denote fits of the form $W(L)=a L+b+c/L+d/L^2$. 
  The insets show derivatives of data with respect to $L$. 
 }\label{figCOMBent}
\end{figure*}

\section{Supplementary Material}

 \subsection{Extracting the disorder strengths $W^*(L)$ and $W_T(L)$ }
 
 The average gap ratio $\left \langle r \right \rangle$ obtained for disordered PXP
 model with the constraint radius $\alpha=2$ is shown in Fig.~\ref{figS1}a). A crossover 
 between the ergodic regime at small disorder strengths $W$ with $\left \langle r \right \rangle\approx \overline r_{GOE} \approx 0.53$
 and the MBL regime at a large disorder 
 with $\left \langle r \right \rangle\approx \overline r_{PS} \approx 0.39$ is clearly visible. 
 
 To extract the disorder strength $W_T(L)$ we find a crossing point of the
 $ \left \langle r \right \rangle (W)$ curve for a given system size $L$ with a constant $\overline r =r_{GOE}-p$,
 as shown in Fig.~\ref{figS1}b). We take $p=0.01$ and we have verified that setting $p \in(0.005,0.02)$ does not change the
 trends for $W_T(L)$ curves reported in the main text.
 
 The disorder strength $W^*(L)$ is obtained as a crossing point of  $ \left \langle r \right \rangle (W)$ curves 
 for system sizes $L_1$, $L_2$ such that $L=(L_1+L_2)/2$. Fig.~\ref{figS1}c) illustrates the extraction 
 of $W^*(L=26)$ for disordered PXP model with $\alpha=2$.

In a similar way one can analyze the crossover between the ergodic and MBL regimes using the average bipartite entanglement entropy 
of eigenstates $\left \langle S  \right \rangle$. The average is performed over $\min\{\mathcal N_{\alpha}/20,1000 \} $ eigenstates in
the middle of the spectrum and over all disorder realizations. Subsequently, the obtained values of the average entanglement 
entropy $\left \langle S  \right \rangle$ are divided by the entanglement entropy of random Gaussian states $S_{RG}$ for the  constrained model 
of radius $\alpha$. The value of $S_{RG}$ is calculated numerically. The resulting rescaled entanglement entropy
$\left \langle S  \right \rangle/S_{RG}$ is expected to be close to $1$ in the ergodic phase and to follow a $1/L$ scaling
at large disorder strengths when the entanglement of eigenstates follows an area-law. This is indeed observed, as we show in 
Fig.~\ref{figEnt}.

The disorder strength $W_T(L)$ is extracted from the rescaled entanglement entropy data at the point 
 at which $\left \langle S  \right \rangle/S_{RG}=0.8$ (other values in the interval $(0.7,0.95)$ give similar results).
The rescaled entanglement entropy $\left \langle S  \right \rangle/S_{RG}$ can be used to determine  crossing points resulting
in $W^*(L)$, analogously as in the case of the average gap ratio. The disorder strengths $W_T(L)$ and $W^*(L)$ obtained  in 
such an analysis of the rescaled entanglement entropy data are shown in Fig.~\ref{figCOMBent}. The linear drift of the 
whole ergodic-MBL crossover with increasing system size $L$ to larger disorder strengths is clearly observed. This
supports the conclusions  obtained from the analysis of the average gap ratio in the main text.

 \subsection{The radius of Fock space}
 
 In this section we provide arguments supporting the expectation that the kinetic term is getting 
 more effective with increasing constraint radius $\alpha$ for the PXP models and that it is 
 not the case for the constraints models with $U(1)$ symmetry. 
 
 Let us take an arbitrary eigenstate $\ket{\psi}$ of $S^z_i$ operators which can be mapped to a
 certain Fock state of spinless fermions.
 Such a state is an eigenstate of the Hamiltonian of the constrained model in the large
 disorder ($W\rightarrow \infty$) limit. We consider now the following procedure.
 Acting with Hamiltonian $H$ on $\ket{\psi}$, we obtain 
 a state $H \ket{\psi}$ which is a non-trivial superposition of Fock states. The minimal integer $R$ for which 
 $H^R \ket{\psi}$ has a non-zero overlap with \emph{all} Fock states in the Hilbert space, averaged
 over initial states, defines a \emph{Fock space radius} $\braket{R}$.
 The bigger the value of $\braket{R}$, the larger is the number of actions of the kinetic term
 of Hamiltonian $H$ needed to reach an arbitrary Fock state. 
 Hence, it may be expected that 
 the localization will be favored in systems where $\braket{R}$ increases rapidly with the system size $L$. 
 And conversely, slow growth of  $\braket{R}$ with system size $L$ means that only few actions of $H$ is
 needed in order to reach an arbitrary state from a given initial state, hence the Fock states are much more prone 
 to delocalization over the whole Hilbert space.
 
\begin{figure}
 \includegraphics[width=0.8\linewidth]{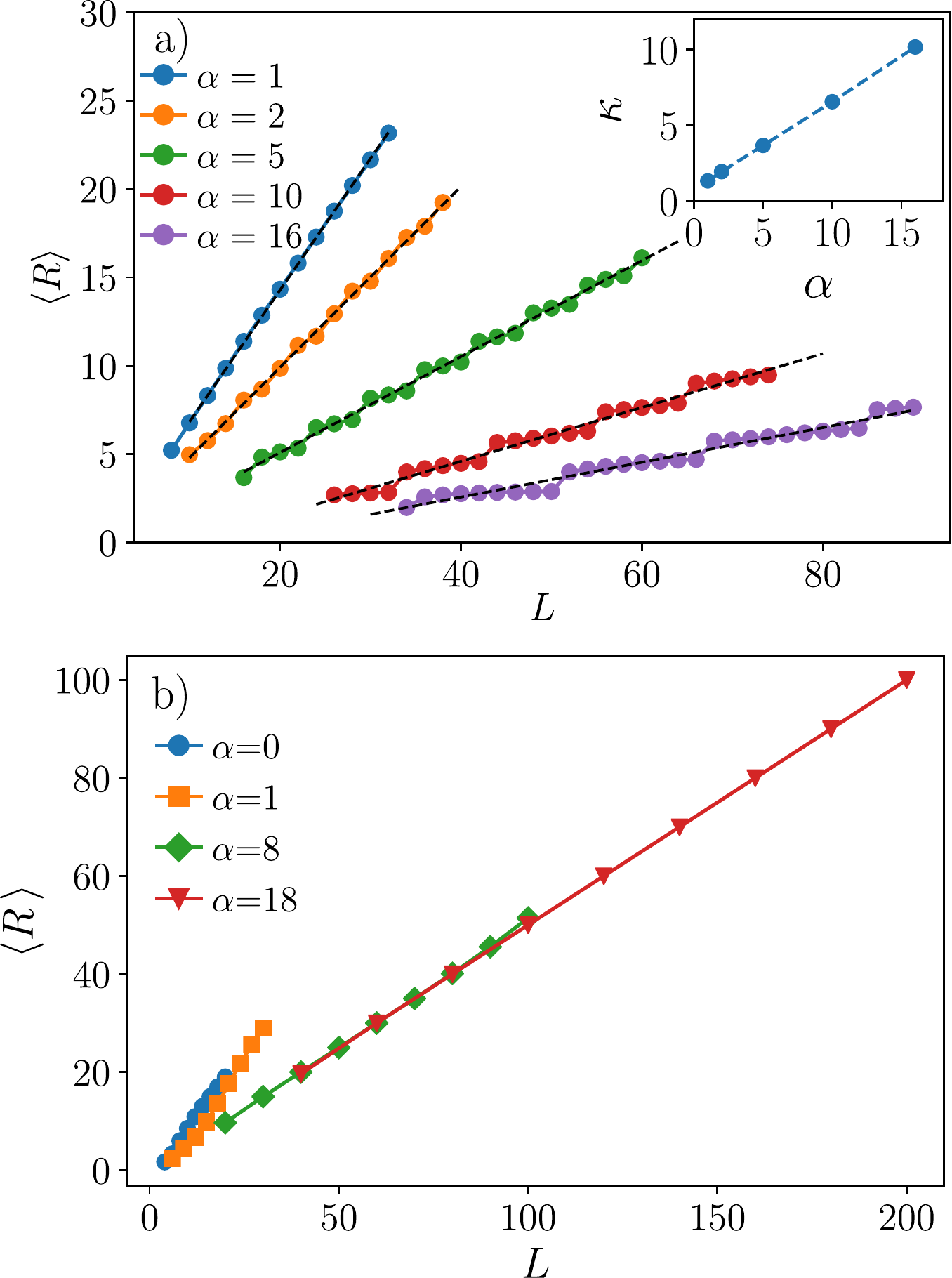} \vspace{-0.3cm}
  \caption{The average radius of the Fock space $\left \langle R  \right \rangle$ as function of the system size $L$
  for varying constrain radius $\alpha$.
  For PXP models -- panel a) -- $\left \langle R  \right \rangle \approx \frac{1}{\kappa} L $ where $\kappa$ is increasing 
  linearly with $\alpha$ (see the inset); for the constrained models with $U(1)$ symmetry, 
  $\left \langle R  \right \rangle \approx \frac{1}{2} L $ for large constraint radius $\alpha$.}
  \label{figRAD}
\end{figure}

 The values of the average Fock space radius $\braket{R}$ for PXP models shown in Fig.~\ref{figRAD} a),
 demonstrate the approximately linear dependence $\braket{R}=\frac{1}{\kappa}L + const$  for all values of constraint
 radius $\alpha$. The coefficient $\kappa$ is linearly increasing with the constrain radius as the 
 inset in Fig.~\ref{figRAD} a) shows. This implies that the disordered PXP models are indeed more
 prone to delocalization as the constraint radius is increased -- this is reflected both in the values of 
 $W^*$ growing with $\alpha$ as well as in the well pronounced linear behavior of the $W^*(L)$ 
 curves at large $L$. In fact, the oscillations on top of the linear trend in $W^*(L)$ for disordered
 PXP models with $\alpha=5, 7, 10$ match the stair-like structure that appears in $\braket{R}(L)$ dependence.
 
 The average Fock space radius $\braket{R}$ for the constrained models with $U(1)$ symmetry
 is shown in Fig.~\ref{figRAD} b) confirming that $\braket{R}\approx \frac{1}{2}L $ at large $\alpha$.
 This makes the constrained models with $U(1)$ symmetry much more prone to localization in comparison to disordered
 PXP models. However, the interplay of disorder and constraints, described in the main text, assures the 
 ergodicity of the constrained models with $U(1)$ symmetry at sufficiently large system size.
 
 \begin{figure}
 \includegraphics[width=0.8\linewidth]{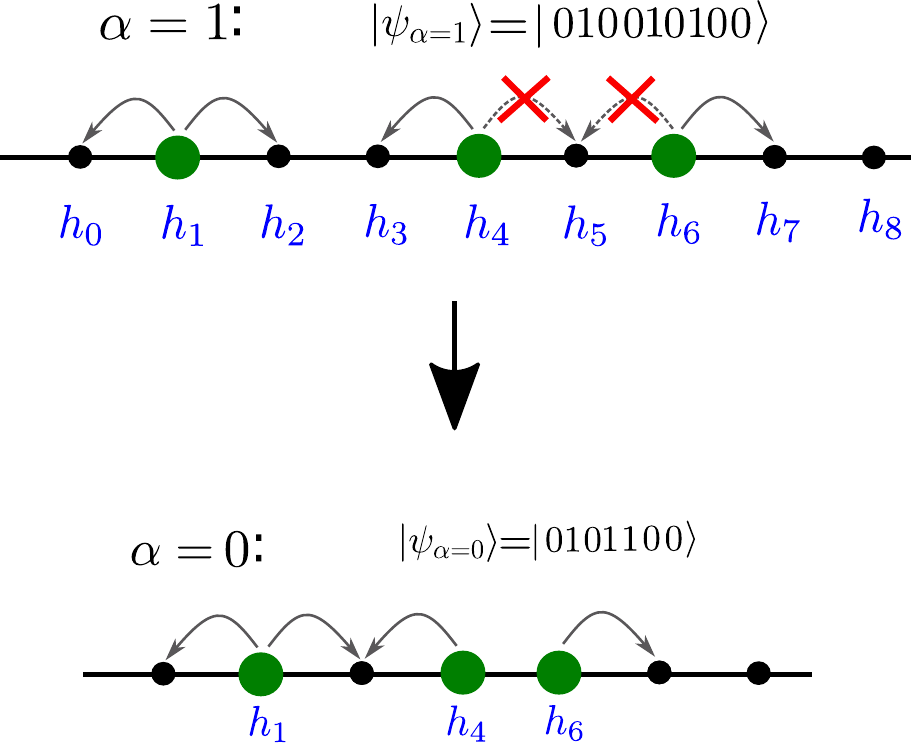} 
  \caption{Mapping between constrained model (constraint radius $\alpha=1$) with $U(1)$ symmetry (top) and an unconstrained model.}
 \label{figMAP}
\end{figure}  
 \subsection{The mapping between constrained and unconstrained models}
 The mapping between constrained and unconstrained models 
 is schematically shown in Fig.~\ref{figMAP}. State $\ket{ \psi_{\alpha=1} }  = \ket{010010100}$ is
 a state of the constrained model with constraint radius $\alpha=1$. Due to the presence of constraints, the particles 
 on sites $i=4$ ($i=6$), where numeration starts from $i=0$, cannot hop to the right (left).
 To map a state of model with constraint radius $\alpha$ and size $L$
 to a state of unconstrained model, each particle is joined with $\alpha$ sites to its right (in some sense
 each particle acquires a $1+\alpha$ site radius). To assure that it can be done for each particle $\alpha$ sites are added
 at the right end of the chain, so that the chain is of size $L+\alpha$. Subsequently, each 'particle' of radius $1+\alpha$ is
 replaced by an ordinary  spinless fermion. The resulting chain size is $L+\alpha - N\alpha$. The procedure can be performed 
 for each Fock state of the constrained model establishing a 1:1 mapping between 
 a constrained model with $N$ particles on $L$ sites with constraint radius $\alpha$ (and open boundary conditions)
 to a model of $N$ spinless fermions on $L-\alpha(N-1)$ sites.

\begin{figure*}
 \includegraphics[width=1\linewidth]{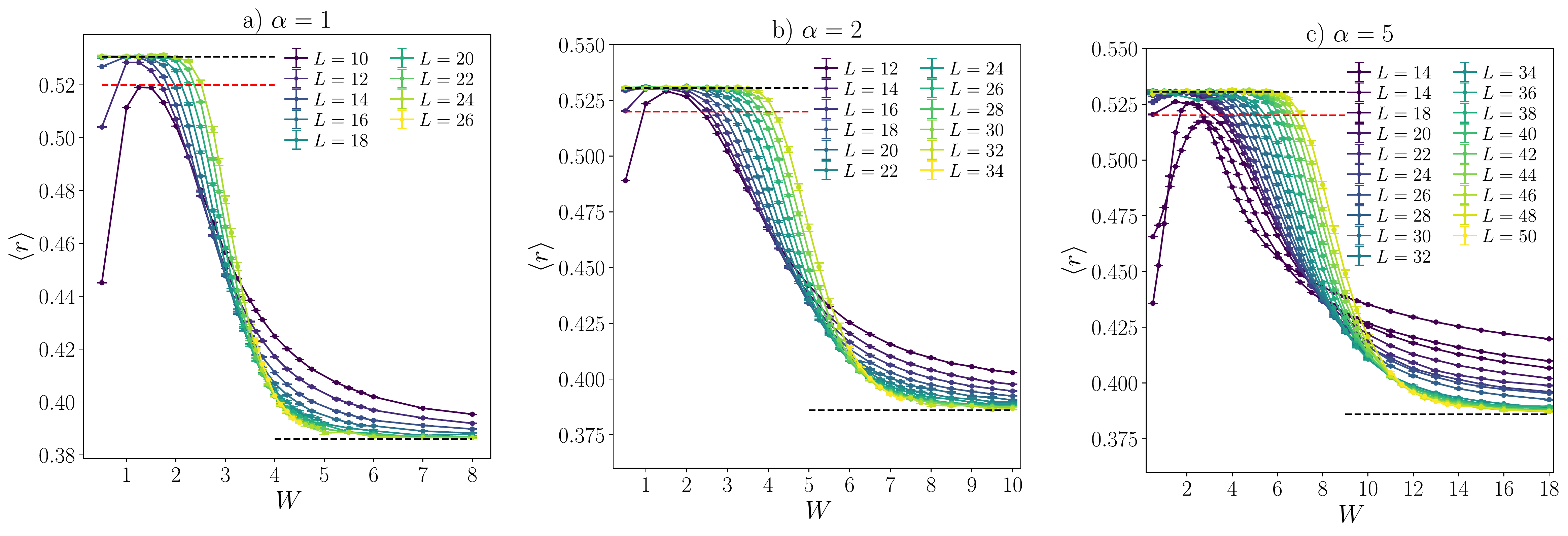}\vspace{-0.3cm}
  \caption{The average gap ratio at the middle of the spectrum of disordered
  PXP models a function of disorder strength $W$ for various 
  system sizes $L$ and radius of constraint $\alpha$.
}\label{figRbar}
\end{figure*}  

\begin{figure*}
 \includegraphics[width=1\linewidth]{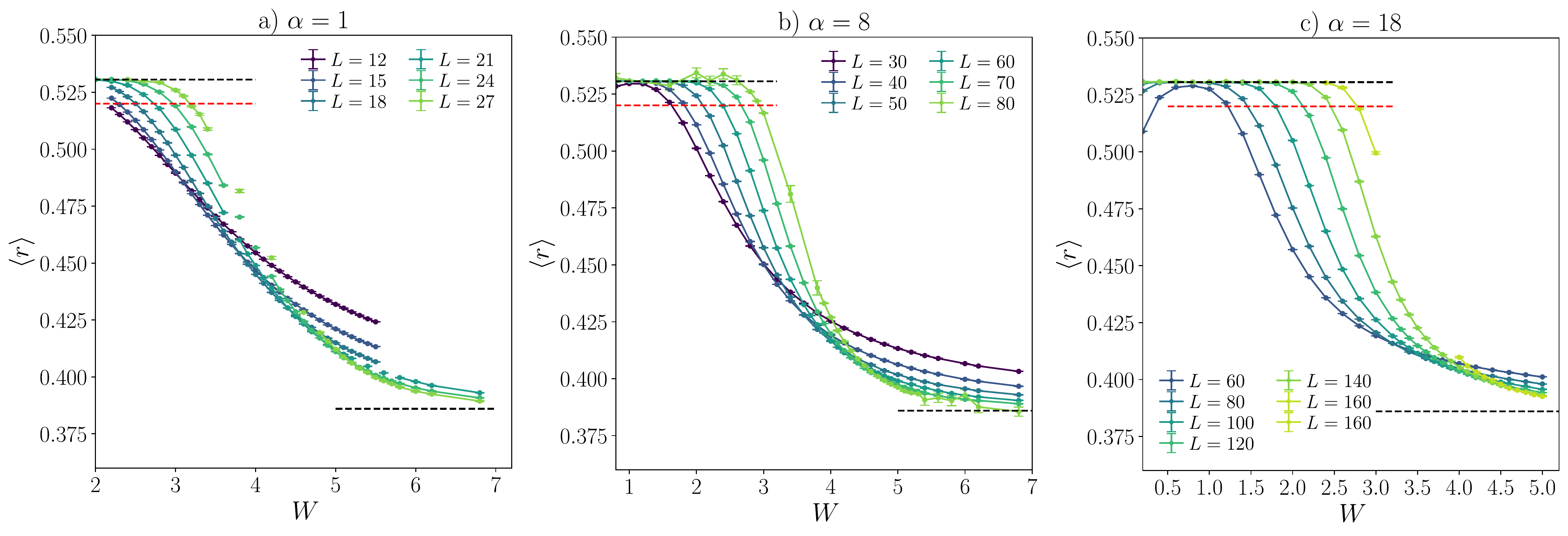}\vspace{-0.3cm}
  \caption{The average gap ratio at the middle of the spectrum of disordered
  constrained models with $U(1)$ symmetry (Hamiltonian (2) in the main text) a function of disorder strength $W$ for various 
  system sizes $L$ and radius of constraint $\alpha$.
}\label{figRbar2}
\end{figure*}  

 \subsection{Average gap ratio for constrained models}
 
 Data for average gap ratio for disordered PXP models and for  constrained models with $U(1)$ symmetry, used to 
 extract the disorder strengths $W^*(L)$ and $W_T(L)$ is shown in Fig.~\ref{figRbar}
 and Fig.~\ref{figRbar2}.

\end{document}